\documentclass[aps,pra,preprint,groupedaddress]{revtex4-1}
\usepackage{graphicx,graphics}
\usepackage[caption=false]{subfig}
\usepackage{amsmath,amssymb}
\usepackage{epstopdf}
\begin{document}

% Use the \preprint command to place your local institutional report
% number in the upper righthand corner of the title page in preprint mode.
% Multiple \preprint commands are allowed.
% Use the 'preprintnumbers' class option to override journal defaults
% to display numbers if necessary
%\preprint{}

%Title of paper
\title{Employing coupled cavities to increase the cooling rate of a levitated nanosphere in the resolved sideband regime}

% repeat the \author .. \affiliation  etc. as needed
% \email, \thanks, \homepage, \altaffiliation all apply to the current
% author. Explanatory text should go in the []'s, actual e-mail
% address or url should go in the {}'s for \email and \homepage.
% Please use the appropriate macro foreach each type of information

% \affiliation command applies to all authors since the last
% \affiliation command. The \affiliation command should follow the
% other information
% \affiliation can be followed by \email, \homepage, \thanks as well.
\author{Mohammad Ali Abbassi}
\email[]{mohammadali.abbassi@ee.sharif.edu}
\author{Khashayar Mehrany}
\email[]{mehrany@sharif.edu}
%\homepage[]{Your web page}
%\thanks{}
%\altaffiliation{}
\affiliation{Department of Electrical Engineering, Sharif University of Technology, Tehran, Iran}

%Collaboration name if desired (requires use of superscriptaddress
%option in \documentclass). \noaffiliation is required (may also be
%used with the \author command).
%\collaboration can be followed by \email, \homepage, \thanks as well.
%\collaboration{}
%\noaffiliation

\date{\today}

\begin{abstract}
In this paper we investigate cooling of a levitated nanosphere in a system of coupled cavities in the resolved sideband regime. Thanks to the presence of an extra resonance in the coupled cavity cooling system, the coupling strength can be maximized at the optimum detuning. In this fashion, the intra-cavity photon number is increased and thereby the cooling rate is enhanced and the strong coupling regime is achieved without resorting to increased driving laser power. The underlying physics of the increased cooling efficiency in the here-proposed system of coupled cavities in the resolved sideband regime and that of the already reported system of coupled cavities in the unresolved sideband regime are significantly different from each other. Since the spectral density of the displacement of the particle can no longer be accurately approximated by the conventional Lorentzian lineshape in the strong coupling regime, a double Lorentzian lineshape is employed to accurately approximate the spectral density of the displacement of the particle and to provide analytical formulations for the cooling rate. The analytical expression given for the cooling rate is validated by numerical simulations.
% insert abstract here
\end{abstract}

% insert suggested PACS numbers in braces on next line
% insert suggested keywords - APS authors don't need to do this
%\keywords{}

%\maketitle must follow title, authors, abstract, \pacs, and \keywords
\maketitle

% body of paper here - Use proper section commands
% References should be done using the \cite, \ref, and \label commands
%\section{}
% Put \label in argument of \section for cross-referencing
%\section{\label{}}
%\subsection{}
%\subsubsection{}
\section{Introduction}
Cooling a mechanical oscillator down to its ground state enables us to realize quantum behaviors by overcoming the thermal noise at mesoscopic and macroscopic scales\cite{Aspelmeyer2014, Marquardt2007, Wilson2008, Genes2008}. Recently, many efforts have been made to realize the mechanical system with a levitated nanosphere which brings forth two-fold benefits. First, the quality factor of the mechanical resonance is much higher when a levitated nanosphere acts as a mechanical oscillator \cite{Chang2010, Romero2010, Yin2011, Yin2013, Romero2011Optically, Romero2011Large, Pflanzer2012}. Second, reaching the strong coupling regime where the mechanical and optical modes are hybridized is facilitated \cite{Monteiro2013}. The nanosphere in such systems can be trapped by an external tweezer \cite{Romero2010}, or by two modes of a single cavity, which is usually referred to as the self trapping scheme\cite{Chang2010, Pender2012}. It is worth noting that two scenarios are conceivable in the self-trapping scheme. The nanosphere can be trapped by one cavity mode and cooled by the other, or it can be trapped and cooled by both cavity modes simultaneously. The former and the latter scenarios are referred to as the single and double resonance schemes, respectively\cite{Pender2012,Monteiro2013}. It is worth noting that the cooling rates in the double resonance scheme are more than one order of magnitude faster than the cooling rates in the single resonance scheme \cite{Pender2012}.\par
The idea of using coupled cavities for cooling in the unresolved sideband regime has drawn much attention in the recent years\cite{Guo2014,Liu2015,Feng2017} since employing the coupled cavities enables us to suppress the heating process thanks to the non-Lorentzian lineshape of the optical resonance in such systems which is either of the Fano (asymmetric) or EIT (symmetric) type.  In this fashion, the ground state cooling can be realized by overcoming the minimum phonon number restriction in the unresolved sideband regime. The impressive success of employing coupled cavities in the unresolved sideband regime begs the question that whether use of the coupled cavities can also be beneficial in the resolved sideband regime. As expected, use of the coupled cavities in the resolved sideband regime proves to be advantageous. It should be however noted that the underlying physics of cooling in the resolved sideband regime is significantly different from the underlying physics of cooling in the unresolved sideband regime. Therefore, the optimum parameters for the coupled cavities in the resolved and unresolved sideband regimes differ from each other and do not lead to the same results. It is later shown that having a non-Lorentzian lineshape for the optical resonance is of no consequence here. Rather, it is important to have the possibility of maximizing the optomechanical coupling strength at the optimum detuning. It is shown that the cooling rate can be considerably enhanced in this fashion and that the strong coupling regime can be reached at much lower driving laser powers.\par 
This paper is organized as follows: in Sec. \ref{sec:TA}, a theoretical analysis of the system is presented in which the Hamiltonian of the system is introduced and the dynamics of the system is studied. In Sec. \ref{sec:OCR} the extraction of the cooling rate is studied, and two closed solutions for the cooling rate are presented in which the spectral density of the displacement of the particle is approximated by a single and double Lorentzian lineshapes. Then, numerical results are presented together with some discussions in Sec. \ref{sec:NR}. Eventually, conclusions are given in Sec. \ref{sec:C}.

\section{\label{sec:TA}Theoretical analysis}
\begin{figure}
\includegraphics[width=0.5\textwidth]{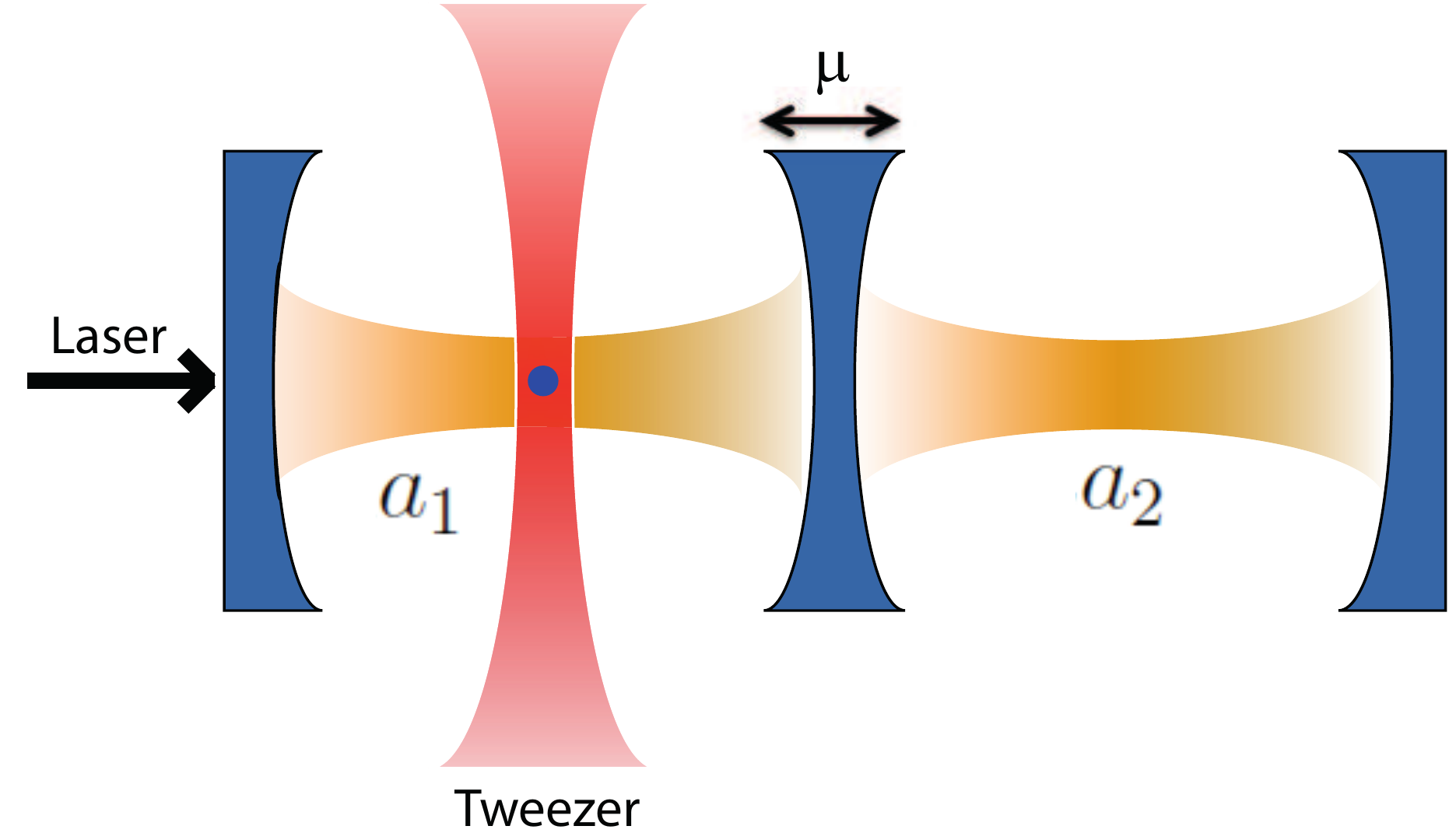}
\caption{\label{fig:sch}Schematics of the arrangement for optical cooling of a nanosphere within coupled cavities.}
\end{figure}

The cooling of a nanosphere in a system of coupled optical cavities is studied in this section. The coupled cavities are schematically shown in Fig. \ref{fig:sch}. The angular frequency and the mode amplitude of the optical cavity in which the nanosphere resides are represented by $\omega_1^{(0)}$ and $a_1$, respectively. This cavity is side-coupled to another cavity whose angular frequency and mode amplitude are represented by $\omega_2^{(0)}$ and $a_2$, respectively. The former is hereafter referred to as the first cavity and the latter as the second cavity. It is assumed that the nanosphere is trapped at position $x_t$ within the first cavity via a trapping laser beam acting as an optical tweezer. The angular frequency of this laser is denoted by $w_t$. The trapped nanosphere is then cooled by a cooling laser whose angular frequency and driving strength are represented by $\omega_L$ and $E$, respectively. It is worth noting that the second cavity is not directly driven. Rather, the mode of the second cavity is excited on account of the electromagnetic coupling between the first and the second cavities whose strength is represented by $\mu$. Since the presence of the nanosphere alters the coupling strength, $\mu$ is in general a function of the trapping position $x_t$.\par
The interaction between the two cavities and the presence of the nanosphere would shift the resonance frequencies of the unperturbed cavities. The change in the frequency of the cavities under the former effect; i.e. interaction between the cavities, are $\omega_1$ and $\omega_2$, and differ from the unperturbed resonance frequencies\cite{Popovic2006}. The change in the frequency of the cavities under the latter effect; i.e. the presence of the nanosphere, is taken into account in the overall Hamiltonian of the system as described in the following sub-section. Given that the nanosphere is trapped outside the second cavity, its effect on the $\omega_2$ is neglected.\par
Below, we first look into the Hamiltonian of the proposed system and then study the dynamics of the system.

\subsection{\label{subsec:H}Hamiltonian of the system}
The overall Hamiltonian of the system is composed of four different terms and can be written down as follows:
\begin{equation}\label{eq:H}H=H_m+H_o+H_{om}+H_d\end{equation}
The first term, $H_m=\frac{p^2}{2m}+\frac{1}{2}m\omega_t^2 x^2$, accounts for the mechanical motion of the nanosphere whose mass, position, and momentum are represented by $m$, $x$, and $p$, respectively. For simplicity's sake, only the motion of the nanosphere along the common axis of both cavities is considered in $H_m$ and thus the problem is solved in the one-dimensional case.\par
The second term, $H_o$, stands for the Hamiltonian of optical modes within the cavities and can be written as:
\begin{equation}H_o=\hbar \omega_1 a_1^\dagger a_1+\hbar \omega_2 a_2^\dagger a_2+\hbar \mu \left(a_1^\dagger a_2+ a_1 a_2^\dagger\right)\end{equation}\par
The third term, $H_{om}$, constitutes the electromagnetic effects of the nanosphere and represents the shift in the resonance frequency of the first cavity induced by the presence of the nanosphere. As already mentioned, the electromagnetic effect of the nanosphere on the second cavity is insignificant and thus $H_{om}$ can be written down as follows:
\begin{equation}H_{om}=-\hbar A \cos^2(k_1 x)a_1^\dagger a_1 \end{equation}
where $A$ is the amplitude of the resonance frequency shift of the first cavity and $\cos^2k_1 x$ is its intensity profile\cite{Neumeier2015}.\par
Eventually,
\begin{equation}H_d=i\hbar E\left(a_1^\dagger e^{-i\omega_L t}-a_1 e^{i\omega_L t}\right)\end{equation}
corresponds to the Hamiltonian of the cooling laser whose angular frequency and driving strength are represented by $\omega_L$ and $E$, respectively. The driving strength is given by $E=\sqrt{\frac{\kappa_{{ex}_1} P}{\hbar \omega_L}}$ where $P$ is the power of the laser, and $\kappa_{{ex}_1}$ is the external decay rate of the first cavity\cite{Romero2010}.\par
\subsection{Dynamics of the system}
The equations of motion corresponding to the overall Hamiltonian given in Eq. (\ref{eq:H}) in the rotating frame of laser are as follows:
\begin{subequations}
\begin{align}\label{eq:eom}\frac{da_1}{dt}&=i\Delta_1a_1+iAa_1\cos^2{k_1x}-i\mu a_2-\frac{\kappa_1}{2}a_1+E\\
\frac{da_2}{dt}&=i\Delta_2a_2-i\mu a_1 -\frac{\kappa_2}{2}a_2\\
\label{eq:eom1}m\frac{d^2x}{dt^2}&=-m\omega_t^2 \left(x-x_t\right)-\hbar k_1 A a_1^\dagger a_1 \sin{2k_1x}-m\Gamma_m\frac{dx}{dt}+\xi\end{align}
\end{subequations}
where $\Delta_j=\omega_L-\omega_j$ is the detuning of the laser from the resonance frequency of the $j^{\mathrm{th}}$ cavity and $\kappa_j$ is its the decay rate which includes both the intrinsic and the extrinsic terms. $\Gamma_m$ is the mechanical damping of the nanosphere, and $\xi$ corresponds to the thermal noise. It can be shown that 
\begin{equation}\langle \xi(t)\xi(t')\rangle=2m\Gamma_m k_B T\delta(t-t')\end{equation}
where $k_B$ is the Boltzmann constant and $T$ is the ambient temperature\cite{Romero2010}.\par
The equilibrium solutions of the system are as follows: 
\begin{subequations}
\label{eq:ss}
\begin{align}\alpha_1&=\frac{E}{\frac{\kappa_1}{2}-i\widetilde{\Delta}_1+\frac{\mu^2}{\frac{\kappa_2}{2}-i\Delta_2}}\\\alpha_2&=\frac{i\mu \alpha_1}{\frac{\kappa_2}{2}-i\Delta_2}\\
x_0&=x_t-\frac{\hbar k_1 A}{m\omega_t^2}|\alpha_1|^2 \sin2k_1x_0\end{align}
\end{subequations}
where $\alpha_1$ and $\alpha_2$ are the steady state values of $a_1$ and $a_2$, respectively and $x_0$ is the equilibrium position of the nanosphere. Furthermore, $\widetilde{\Delta}_1=\Delta_1+A\cos^2(k_1 x_0)$ is the modified detuning of the first cavity due to the presence of the nanosphere. If the power of the trapping laser is much greater than the power of the cavity mode, then $x_0$ will be almost equal to $x_t$. Otherwise, the set of Eqs. (\ref{eq:ss}) should be solved numerically to find the equilibrium solutions of the system.
\begin{figure}
\includegraphics[width=0.5\textwidth]{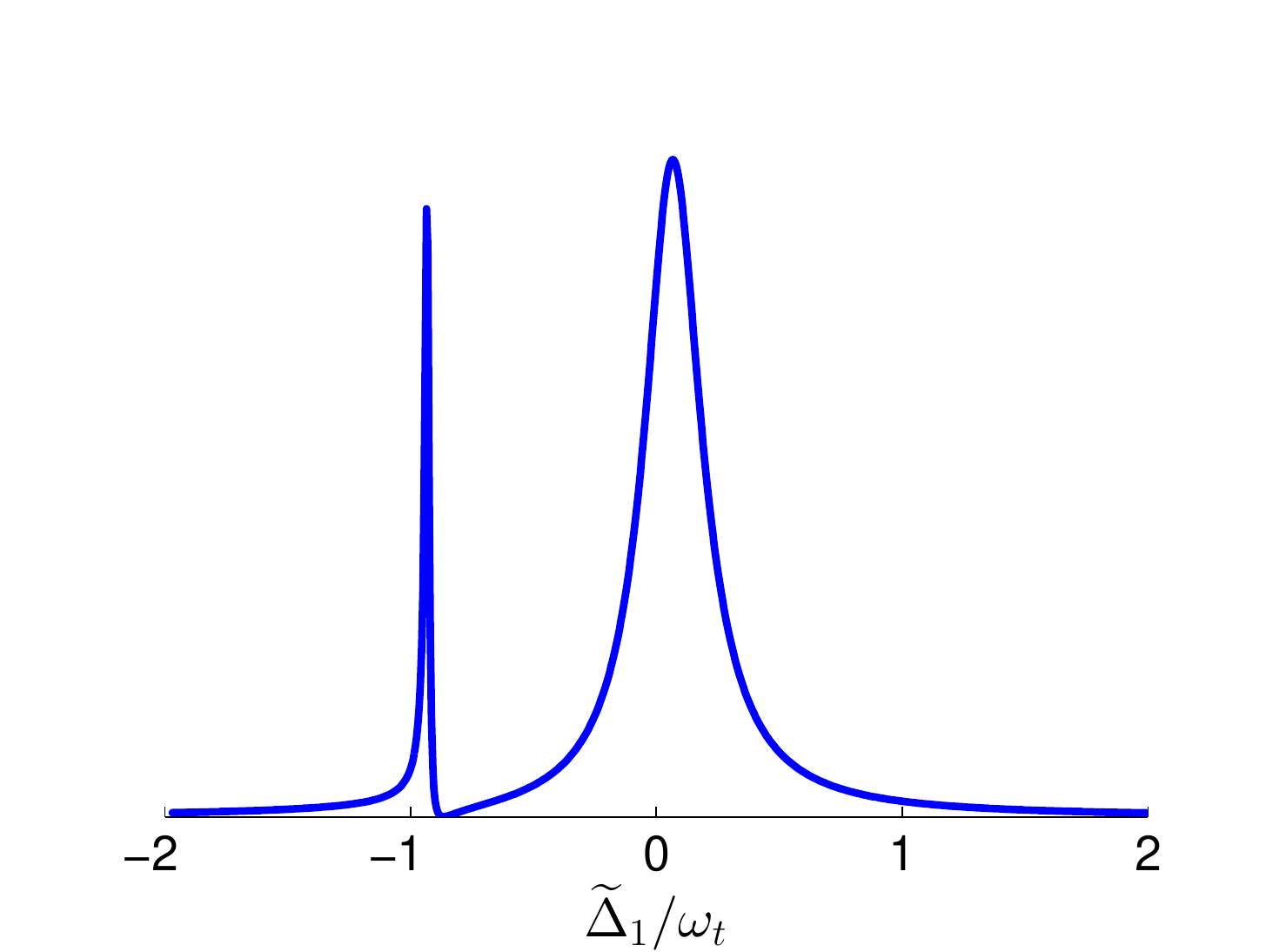}
\caption{The number of photons inside the first cavity.}
\label{fig:OR}
\end{figure}
Fig. \ref{fig:OR} schematically shows the steady state values of the photon number inside the first cavity ($|\alpha_1|^2$) with respect to $\widetilde{\Delta}_1$. The photon number inside the first cavity shows the typical Fano resonance behavior.\par
Now, we can linearize the equations of motion by considering small fluctuations near the equilibrium solutions and obtain the following linear equations by neglecting higher order terms:
\begin{subequations}
\begin{align}\frac{da_1}{dt}&=\left(i\widetilde{\Delta}_1-\frac{\kappa_1}{2}\right)a_1-i\mu a_2-ig_0\alpha_1 x\\
\frac{da_2}{dt}&=\left(i\Delta_2-\frac{\kappa_2}{2}\right)a_2-i\mu a_1\\
m\frac{d^2 x}{dt^2}&=-m\omega_m^2 x -\hbar g_0(\alpha_1^\ast a_1+\alpha_1 a_1^\dagger)-m\Gamma_m\frac{dx}{dt}+\xi\end{align}
\end{subequations}
where $\omega_m$ is the mechanical resonance frequency given by
\begin{equation}\omega_m^2=\omega_t^2+\frac{2\hbar k_1^2 A}{m}|\alpha_1|^2\cos(2k_1x_0)\end{equation}
and
\begin{equation}g_0=k_1A\sin(2k_1x_0)\end{equation}
It can be shown that, the spectral density of $x$, when the system is stable, is given by
\begin{equation}\label{eq:Sx}S_{xx}(\omega)=\frac{1}{2\pi}\int\langle x^\ast(\omega')x(\omega)\rangle d\omega'=2m\Gamma_mk_BT|\chi(\omega)|^2\end{equation}
where
\begin{equation}\chi(\omega)=\frac{x(\omega)}{\xi(\omega)}=\frac{1}{m\left(\omega_m^2-\omega^2-i\omega\Gamma_m\right)-2i m\omega_m g^2\big(\chi_o(\omega)-\chi_o^\ast(-\omega)\big)}\end{equation}
is the mechanical response of the system. Furthermore, $g= g_0|\alpha_1|\sqrt{\frac{\hbar}{2m\omega_m}}$ is the optomechanical coupling strength, and
\begin{equation}\chi_o(\omega)=\frac{a(\omega)}{x(\omega)}=\frac{1}{\frac{\kappa_1}{2}-i\left(\omega+\widetilde{\Delta}_1\right)+\frac{\mu^2}{\frac{\kappa_2}{2}-i\left(\omega+\Delta_2\right)}}\end{equation}
is the optomechanical response of the system.\par
\begin{figure}
\subfloat[\label{fig:Sxx_a}]{\includegraphics[width=0.4\textwidth]{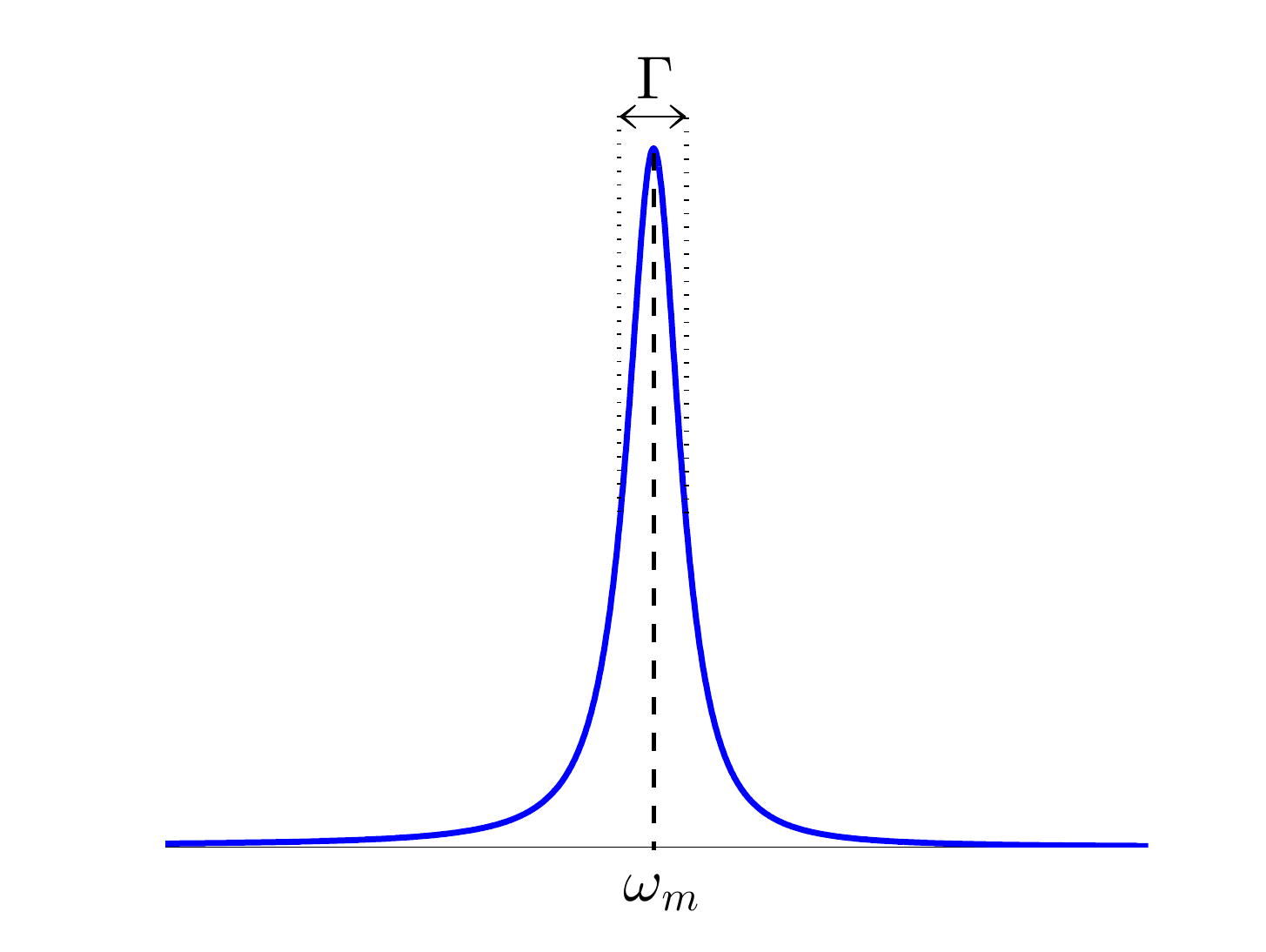}}
\subfloat[\label{fig:Sxx_b}]{\includegraphics[width=0.4\textwidth]{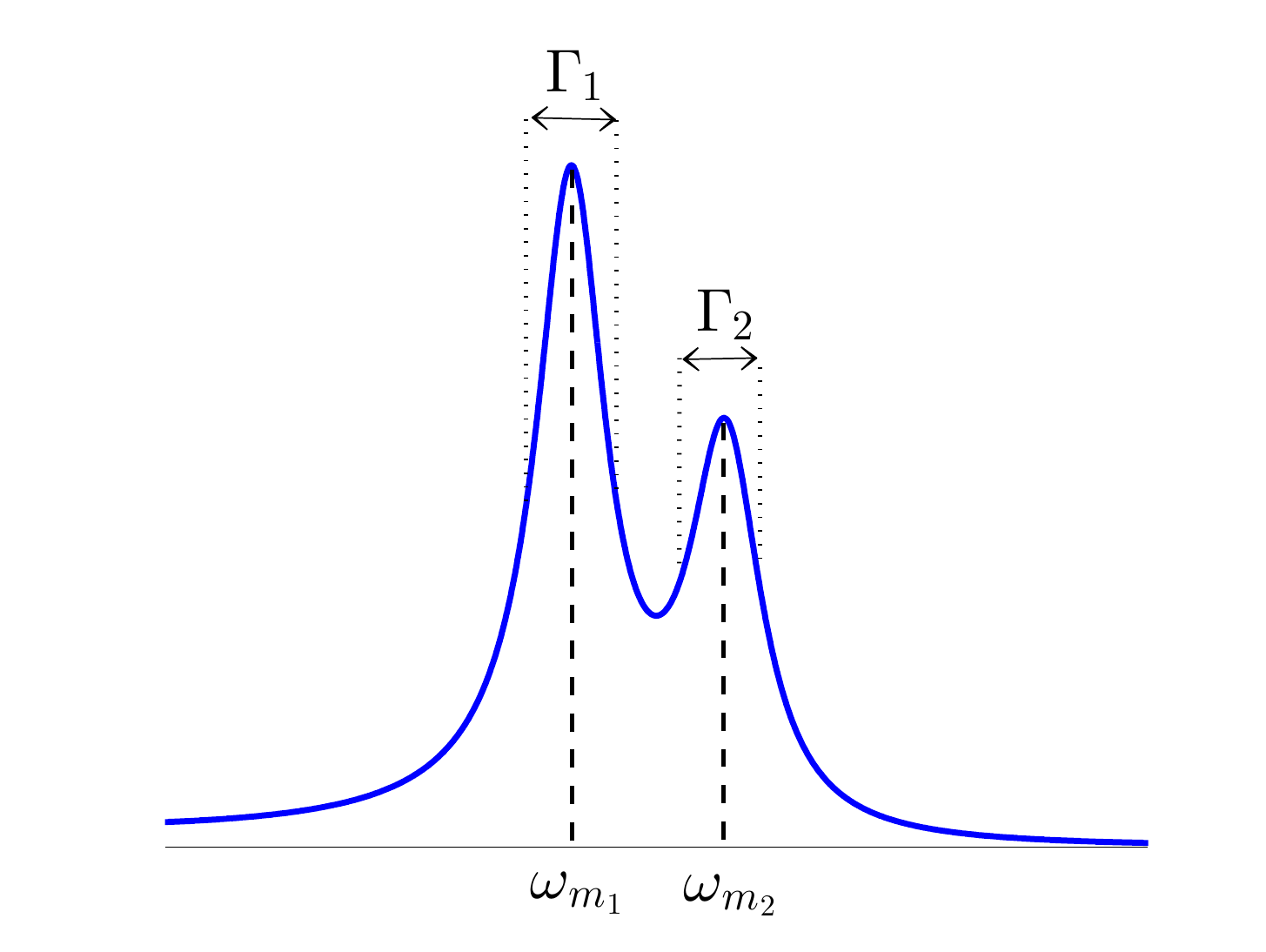}}
\caption{Spectral density of $x$ in (a) weak coupling regime (b) strong coupling regime. }
\label{fig:Sxx}
\end{figure}
It should be noted that in the weak coupling regime where $g\ll \kappa_1$, the spectral density of $x$ can be accurately approximated by a single Lorentzian lineshape around the resonance frequency $\omega_m$:
\begin{equation}\label{eq:SxxSL}S_{xx}(\omega) \approx \frac{2m^{-1}\Gamma_m k_B T}{\left(\omega^2-\omega_m^2\right)^2+\left(\omega\Gamma\right)^2}\end{equation}
This is schematically demonstrated in Fig. \ref{fig:Sxx_a}. This approximation is not valid at the strong coupling regime when $g\sim \kappa_1$. In this regime, however, $S_{xx}$ is made of two distinct resonances caused by the hybridization of the optical and mechanical modes, and thus can be approximated by a double Lorentzian lineshape as
\begin{equation}\label{eq:SxxDL}S_{xx}(\omega)\approx\frac{C_1}{\left(\omega^2-\omega_{m_1}^2\right)^2+\left(\omega\Gamma_1\right)^2}+\frac{C_2}{\left(\omega^2-\omega_{m_2}^2\right)^2+\left(\omega\Gamma_2\right)^2}\end{equation}
where $\omega_{m_{1,2}}$ and $\Gamma_{1,2}$ are the resonance frequencies and decay rates of the nanosphere, respectively, and $C_1$ and $C_2$ are constants to be determined later. This is schematically demonstrated in Fig. \ref{fig:Sxx_b}.
\section{\label{sec:OCR}optical cooling rate}
The optical cooling rate can be obtained by comparing the averaged mechanical energy, $E_m=\frac{\langle p^2\rangle}{2m}+\frac{1}{2}m\omega_m^2\langle x^2\rangle$, against $k_BT$, where $k_B$ is the Boltzmann constant and $T$ is the ambient temperature. The thermal motion of the nanosphere is optically cooled when the averaged mechanical energy is less than $k_BT$. In fact, it is already shown that 
\begin{equation}\label{eq:Em1}E_m=\frac{\Gamma_m}{\Gamma_m+\Gamma_{opt}}k_BT\end{equation}
where $\Gamma_{opt}$ is the sought-after optical cooling rate. The effective mechanical damping rate is defined as $\Gamma_{eff}=\Gamma_m+\Gamma_{opt}$\cite{Aspelmeyer2014}.\par
Now, the averaged mechanical energy, $E_m$, can be written in terms of the spectral density of $x$ by calculating the averaged mechanical energy in the frequency domain:
\begin{equation}\label{eq:Em2}E_m=\frac{1}{4\pi}\int m\left(\omega_m^2+\omega^2\right)S_{xx}(\omega) d\omega\end{equation}
Using Eq. (\ref{eq:Sx}) in comparing Eq. (\ref{eq:Em1}) and Eq. (\ref{eq:Em2}) yields the effective mechanical damping rate:
\begin{equation}\label{eq:Geff}\Gamma_{eff}=\frac{2\pi}{m^2\int(\omega_m^2+\omega^2)|\chi(\omega)|^2d\omega}\end{equation}
It is worth noting that the mechanical motion of the particle is optically cooled when $\Gamma_{eff} > \Gamma_m$, i.e. when $\Gamma_{opt} > 0$. In case $\Gamma_{eff}<\Gamma_{m}$, i.e. when $\Gamma_{opt} < 0$, the mechanical motion is in fact optically heated. Now, if the effective mechanical damping rate becomes negative then the system is unstable and the expression in Eq. (\ref{eq:Geff}) is no longer valid.\par
To obtain analytical expressions for the optical cooling rate, the single and double Lorentzian approximation of the spectral density of $x$ given in Eq. (\ref{eq:SxxSL}) and Eq. (\ref{eq:SxxDL}) are employed for the weak and strong coupling regimes, respectively.
\subsection{Single Lorentzian approximation} 
As already mentioned, the mechanical response of the system can be approximated by a single Lorentzian lineshape with the resonance frequency $\omega_m$ in the weak coupling regime. Thus, the optical cooling rate is given by
\begin{equation}\Gamma_{opt}=2g^2\mathrm{Re}\big[\chi_o(\omega_m)-\chi_o^\ast(-\omega_m)\big]\end{equation}
which can be further simplified to
\begin{equation}\label{eq:G0}\Gamma_{opt}=\frac{g^2\kappa_1}{\left(\frac{\kappa_1}{2}\right)^2+\left(\omega_m+\widetilde{\Delta}_1-\frac{\mu^2}{\omega_m+\Delta_2}\right)^2}-\frac{g^2\kappa_1}{\left(\frac{\kappa_1}{2}\right)^2+\left(\omega_m-\widetilde{\Delta}_1-\frac{\mu^2}{\omega_m-\Delta_2}\right)^2}\end{equation}
when the decay rate of the second cavity is almost negligible ($\kappa_2 \approx 0$). The first and second term in the above expression are the rates of the anti-Stokes (cooling) and Stokes (heating) processes, respectively. When $\widetilde{\Delta}_1<0$, the cooling process is dominant, and we can ignore the second term in Eq. (\ref{eq:G0}). The denominator of the first term is minimized at
\begin{equation}\label{eq:SLC1}\widetilde{\Delta}_1=-\omega_m-\frac{d}{2}\pm \sqrt{\frac{d^2}{4}+\mu^2}\end{equation} 
where $d=\Delta_2-\widetilde{\Delta}_1$, while its numerator, which is proportional to the number of photons inside the first cavity ($|\alpha_1|^2$) is maximized at
\begin{equation}\label{eq:SLC2}\widetilde{\Delta}_1=-\frac{d}{2}\pm \sqrt{\frac{d^2}{4}+\mu^2}\end{equation}
In the latter expression, the positive and negative signs correspond to the detuning values which satisfy Lorentzian and Fano resonance conditions, respectively. It can be shown that Eq. (\ref{eq:SLC1}) and Eq. (\ref{eq:SLC2}) are both satisfied when
 \begin{equation}\label{eq:SLC3}\omega_m=\sqrt{d^2+4\mu^2}\end{equation}
Therefore, the cooling process is almost maximized at $\widetilde{\Delta}_{1}=-\frac{\omega_m+d}{2}$. It should be noted that the possibility of maximizing the numerator while the denominator is minimized is achieved as a result of the presence of an extra Fano resonance in the spectrum of $|\alpha_1|^2$, and it is impossible to optimize both the denominator and numerator of the cooling process when the cavities are decoupled ($\mu=0$).
\subsection{Double Lorentzian approximation}
As seen in Fig. \ref{fig:Sxx_b}, the mechanical response of the system cannot be approximated by a single Lorentzian lineshape at the strong coupling regime where $g\sim \kappa_1$. In this regime, $S_{xx}$ is made of two distinct resonances and should be approximated by a double Lorentzian lineshape as given in Eq. (\ref{eq:SxxDL}). Inserting this approximated function in Eq. (\ref{eq:Geff}) and computing the integral by complex analysis yields
\begin{equation}\label{eq:Gamma_DR}\frac{1}{\Gamma_{eff}}=\frac{C_1\left(\omega_m^2+\omega_{m_1}^2\right)}{2\omega_{m_1}^2\Gamma_1}+\frac{C_2\left(\omega_m^2+\omega_{m_2}^2\right)}{2\omega_{m_2}^2\Gamma_2}\end{equation}
Now, we should obtain $\omega_{m_{1,2}}$ and $\Gamma_{1,2}$ as well as the constants $C_1$ and $C_2$. When $\omega_{m_{1,2}}$ and $\Gamma_{1,2}$ are found, we can extract the constants $C_1$ and $C_2$ from exact values of $S_{xx}(\omega_{m_{1,2}})$ given in Eq. (\ref{eq:Sx}) as follows:  
\begin{equation}C_1=\frac{2m\Gamma_m k_BT}{\mathcal{D}}\left(\frac{|\chi(\omega_{m_1})|^2}{\omega_{m_1}^2\Gamma_1^2}+\frac{|\chi(\omega_{m_2})|^2}{\left(\omega_{m_1}^2-\omega_{m_2}^2\right)^2 +\omega_{m_1}^2\Gamma_2^2}\right)\end{equation}
\begin{equation}C_2=\frac{2m\Gamma_m k_BT}{\mathcal{D}}\left(\frac{|\chi(\omega_{m_1})|^2}{\left(\omega_{m_1}^2-\omega_{m_2}^2\right)^2 +\omega_{m_2}^2\Gamma_1^2}+\frac{|\chi(\omega_{m_2})|^2}{\omega_{m_2}^2\Gamma_2^2}\right)\end{equation}
\begin{equation}\mathcal{D}=\frac{1}{\left(\omega_{m_1}\omega_{m_2}\Gamma_1\Gamma_2\right)^2}- \frac{1}{\left(\left(\omega_{m_1}^2-\omega_{m_2}^2\right)^2 +\omega_{m_1}^2\Gamma_2^2\right)\left(\left(\omega_{m_1}^2-\omega_{m_2}^2\right)^2 +\omega_{m_2}^2\Gamma_1^2\right)}\end{equation}
\par
We can find $\omega_{m_{1,2}}$ and $\Gamma_{1,2}$ from the natural frequencies of the system. The natural frequencies of the system are the roots of characteristic polynomial that is given by
\begin{equation}P(s)=m\left(s^2+\Gamma_m s+\omega_m^2\right)R(s)+T(s)\end{equation}
where
\begin{equation}R(s)=\left(\left(\frac{\kappa_1}{2}+s\right)\left(\frac{\kappa_2}{2}+s\right)-\widetilde{\Delta}_1\Delta_2+\mu^2\right)^2+\left(\widetilde{\Delta}_1\left(\frac{\kappa_2}{2}+s\right)+\Delta_2\left(\frac{\kappa_1}{2}+s\right)\right)^2\end{equation}
\begin{equation}T(s)=4m\omega_m g^2\left(\widetilde{\Delta}_1\left(\frac{\kappa_2}
{2}+s\right)^2+\widetilde{\Delta}_1\Delta_2^2-\Delta_2\mu^2\right)\end{equation}
We can expand $P(s)$ around
$s_0=- i\omega_m-\frac{\Gamma_m}{2}$ up to the second order term that leads to
\begin{equation}P(s)=P(s_0)+P'(s_0)\left(s-s_0\right)+\frac{1}{2}P''(s_0)\left(s-s_0\right)^2\end{equation}
Hence, the natural frequencies of the system can be estimated as
\begin{equation}s_{1,2}=s_0+\frac{-P'(s_0)\pm \sqrt{{P'}^2(s_0)-2P''(s_0)P(s_0)}}{P''(s_0)}\end{equation}
and consequently, the resonance frequencies and decay rates of the system are given by
\begin{equation}\omega_{m_{1,2}}=\operatorname{Im}\left[s_0+\frac{P'(s_0)\pm \sqrt{{P'}^2(s_0)-2P''(s_0)P(s_0)}}{P''(s_0)}\right]\end{equation}
\begin{equation}\Gamma_{1,2}=2 \operatorname{Re}\left[s_0+\frac{P'(s_0)\pm \sqrt{{P'}^2(s_0)-2P''(s_0)P(s_0)}}{P''(s_0)}\right]\end{equation}
Now, the effective cooling rate can be obtained by inserting these quantities in Eq. \ref{eq:Gamma_DR}. Some numerical results are given in the next section.
\section{\label{sec:NR} Numerical Results}
As a numerical example, consider a nanosphere with mass $m\mathrm{=9.2\times 10^{-18}[Kg]}$ which is trapped in a system of coupled cavities. The decay rates of the cavities are considered as $\kappa_1\mathrm{=6\times 10^5[Hz]}$ and $\kappa_2\mathrm{=10^3[Hz]}$  ,respectively. The trapping frequency of the laser is considered as $\omega_t\mathrm{=2[MHz]}$ and the trapping position, $x_t$, is adjusted so that $\cos(2k_1x_0)=0$ and consequently $\omega_m=\omega_t$. Other parameters are considered as: $A\mathrm{=10^5 [Hz]}$, $k_1\mathrm{=3\times 10^6 [m^{-1}]}$, and $\Gamma_m\mathrm{=10^{-3} [Hz]}$.\par
\begin{figure}[ht]
\subfloat[Single Lorentzian]{\includegraphics[width=0.33\textwidth]{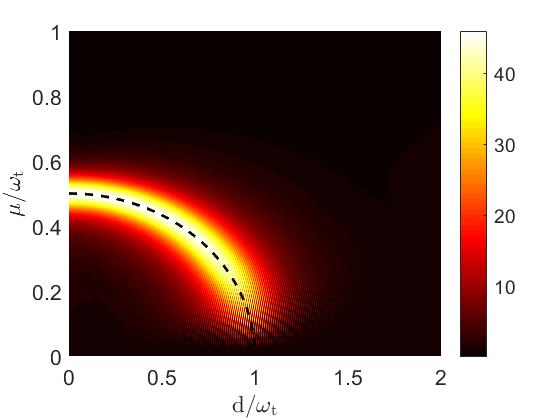}}
\subfloat[Double Lorentzian]{\includegraphics[width=0.33\textwidth]{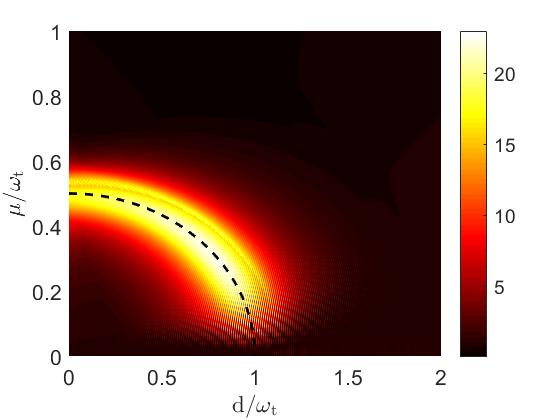}}
\subfloat[Exact solution]{\includegraphics[width=0.33\textwidth]{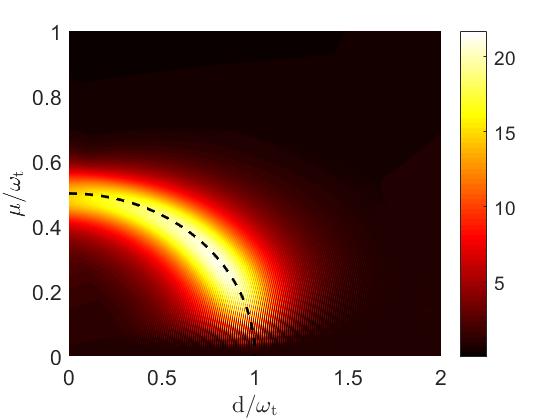}}
\caption{Maximum achievable cooling rate in the plane of $\mu$-$d$. The results are normalized to the maximum cooling rate of a single cavity.}
\label{fig:CR_mu_d}
\end{figure}
\begin{figure}[ht]
\subfloat[Single Lorentzian]{\includegraphics[width=0.33\textwidth]{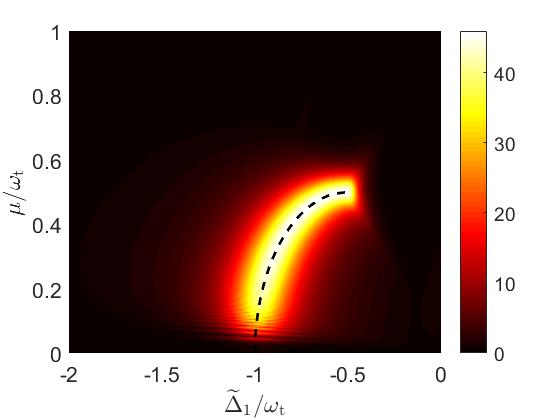}}
\subfloat[Double Lorentzian]{\includegraphics[width=0.33\textwidth]{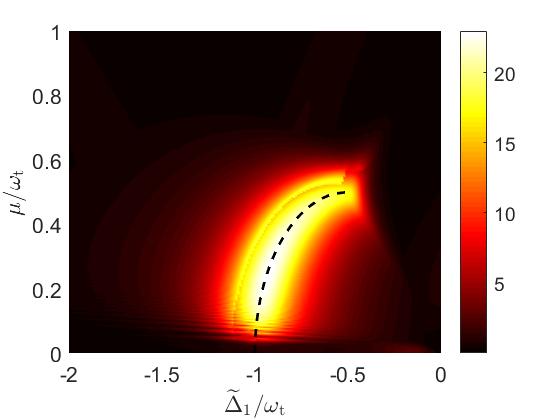}}
\subfloat[Exact solution]{\includegraphics[width=0.33\textwidth]{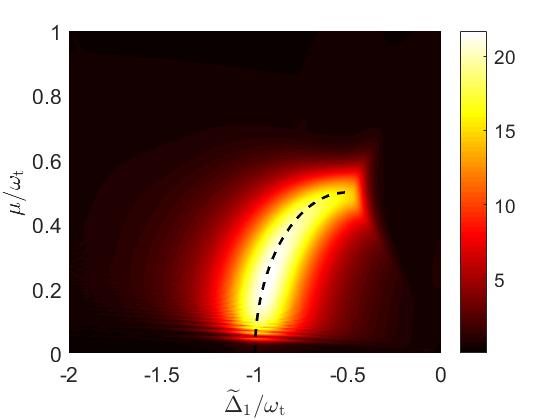}}
\caption{Maximum achievable cooling rate in the plane of $\widetilde{\Delta}_1$-$\mu$. The results are normalized to the maximum cooling rate of a single cavity.}
\label{fig:CR_Delta_mu}
\end{figure}
\begin{figure}[ht]
\subfloat[Single Lorentzian]{\includegraphics[width=0.33\textwidth]{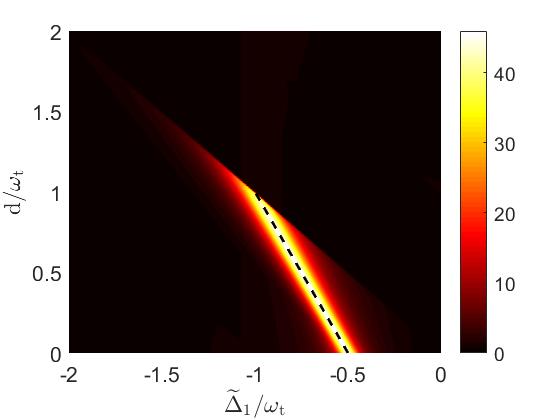}}
\subfloat[Double Lorentzian]{\includegraphics[width=0.33\textwidth]{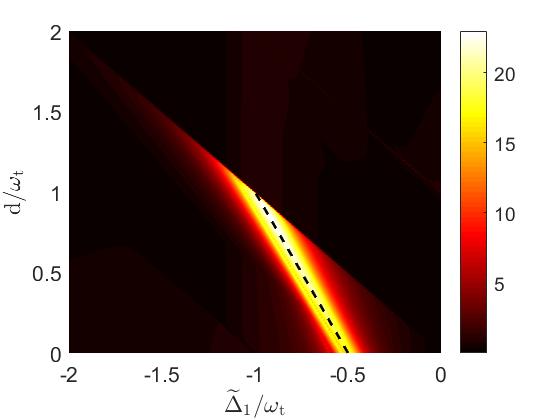}}
\subfloat[Exact solution]{\includegraphics[width=0.33\textwidth]{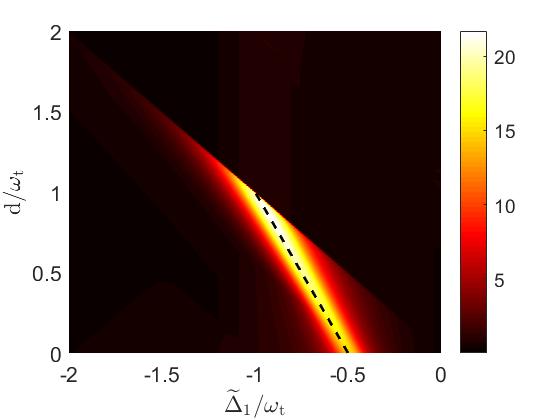}}
\caption{Maximum achievable cooling in the plane of $\widetilde{\Delta}_1$-$d$. The results are normalized to the maximum cooling rate of a single cavity.}
\label{fig:CR_Delta_d}
\end{figure}

First, we investigate how the variation of $\widetilde{\Delta}_1$, $d$, and $\mu$ in the system of coupled cavities can influence the behavior of the cooling rate. To this end, the maximum achievable cooling rate is plotted versus normalized parameters $\mu / \omega_t$ and $d / \omega_t$ in Fig. \ref{fig:CR_mu_d}. The maximum achievable cooling rate can be obtained by finding the maximum cooling rate with respect to $\widetilde{\Delta}_1$ for each values of $\mu$ and $d$. Similarly, the maximum achievable cooling rate in the planes of $\widetilde{\Delta}_1$-$\mu$ and $\widetilde{\Delta}_1$-$d$ are shown in Fig. \ref{fig:CR_Delta_mu} and Fig. \ref{fig:CR_Delta_d}, respectively. In these figures, the cooling rates are calculated by three different methods: single Lorentzian approximation, double Lorentzian approximation, and the exact solution obtained from the calculation of the integral in Eq. (\ref{eq:Geff}) numerically. The dashed lines indicate the approximated contours along which the cooling rate is maximized obtained from Eq. (\ref{eq:SLC3}) when the single Lorentzian approximation is valid.\par
Close inspection of these figures reveals the following facts: albeit the single Lorentzian approximation fails in the calculation of the cooling rates accurately, as a result of reaching the strong coupling regime, it surprisingly predicts the conditions for achieving optimum cooling properly. However, the results of double Lorentzian approximation are almost the same with the exact solutions. Eventually, it is worth noting that more than one order of magnitude faster cooling rates can be achieved when cavities are coupled in comparison with a single cavity that indicates enhancing the efficiency of cooling.\par
\begin{figure}
\subfloat[\label{fig:P_a}]{\includegraphics[width=0.4\textwidth]{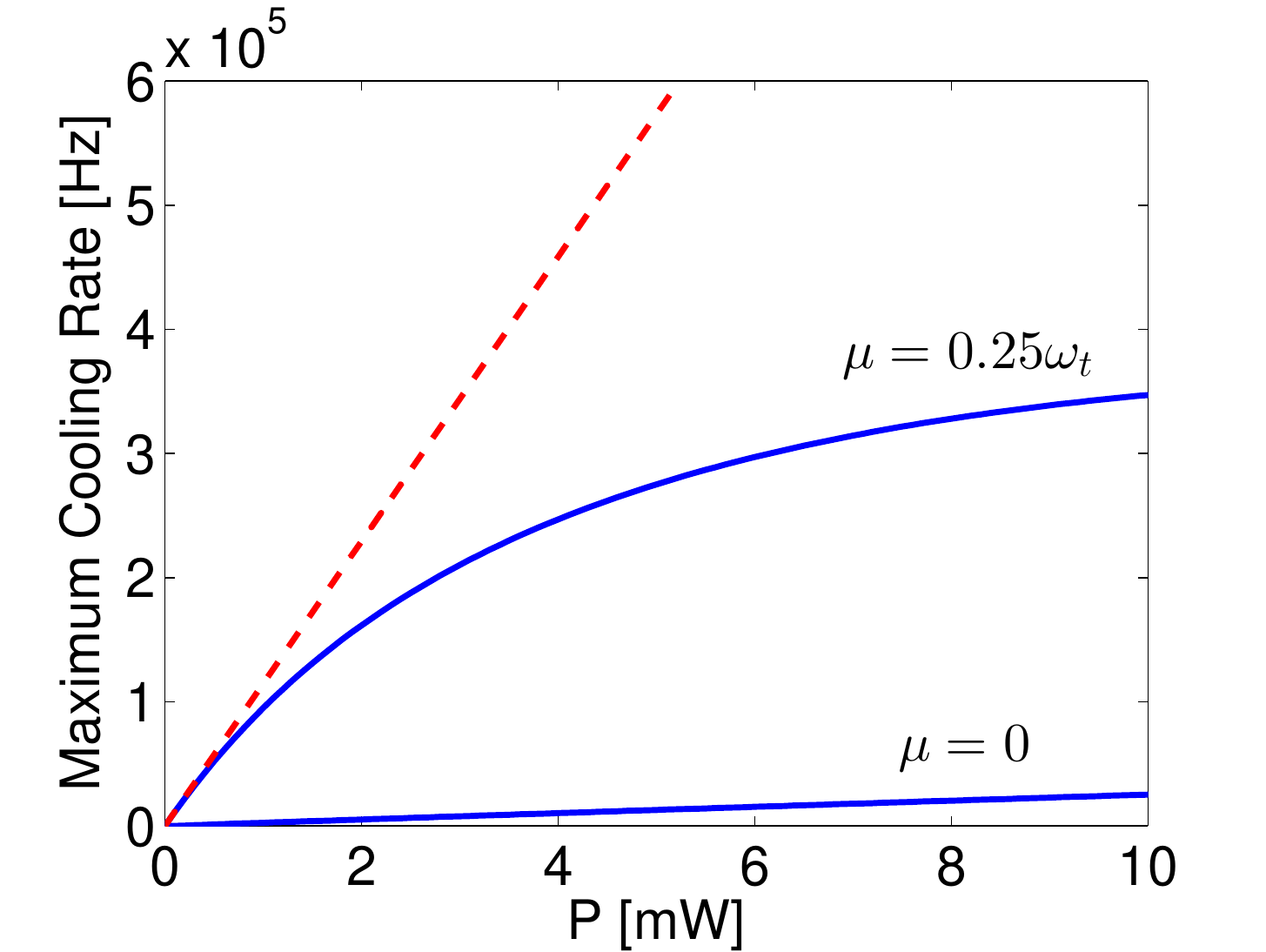}}
\subfloat[\label{fig:P_b}]{\includegraphics[width=0.4\textwidth]{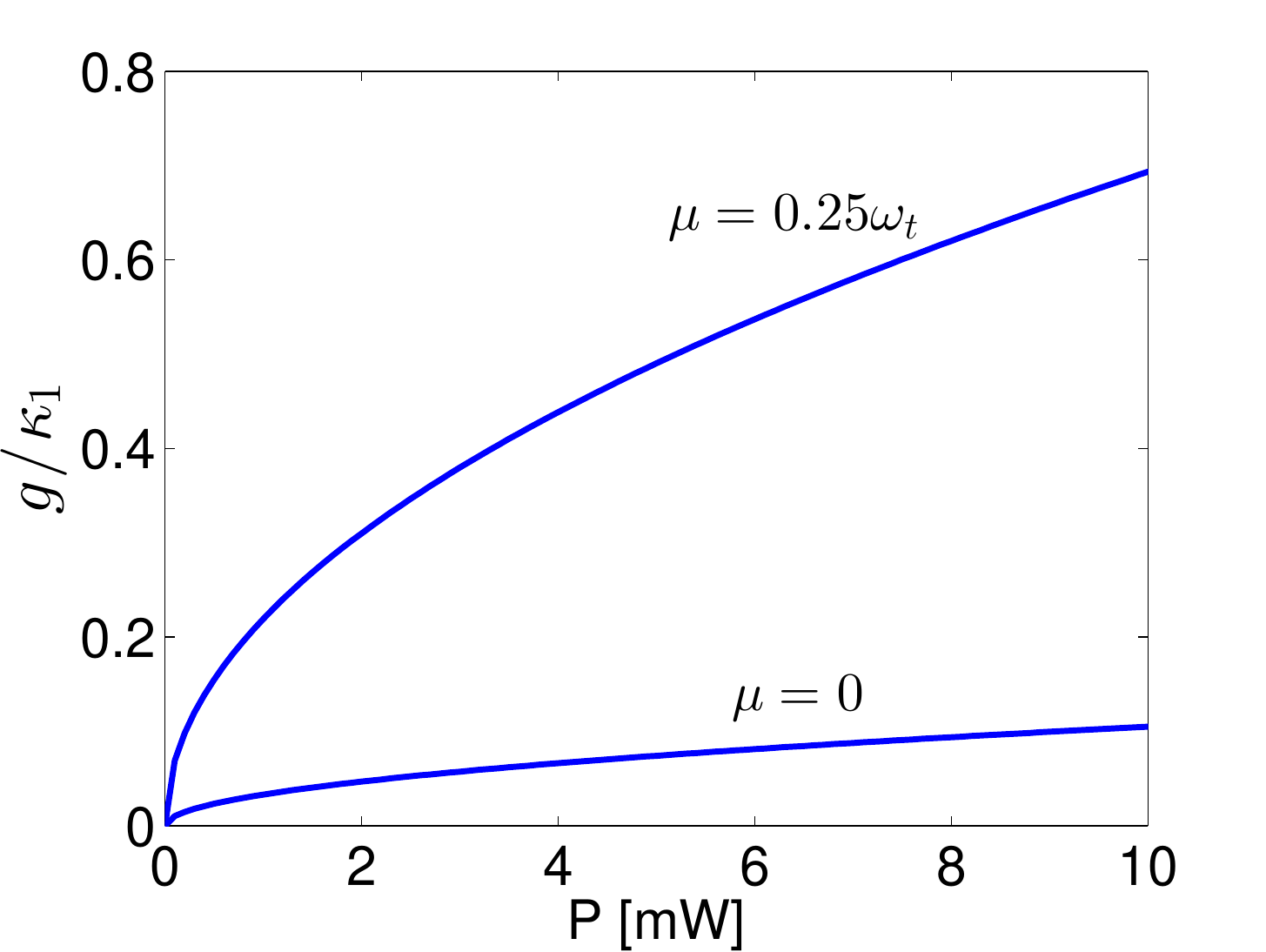}}
\caption{\label{fig:P}(a) Maximum cooling rate versus the laser power. The dashed line shows the results of the single Lorentzian approximation for the case of $\mu=0.25\omega_t$. (b) Optomechanical coupling strength normalized to the decay rate of the first cavity versus the laser power.}
\end{figure}

In the former simulations, the power of the cooling laser was adjusted at $P=\mathrm{5[mW]}$. Now, we study how the power of the cooling laser can affect the optomechanical behavior of the system. The study is carried out for both coupled and decoupled cavities. The maximum cooling rate at the optimum cooling condition given in Eq. (\ref{eq:SLC3}) is shown in Fig. \ref{fig:P_a} for the both cases of the coupled ($\mu=0.25\omega_t$) and decoupled ($\mu=0$) cavities. According to this figure, when the cavities are decoupled, the maximum cooling rate grows linearly in the practical power ranges. However, it can become saturated even at the conventional powers when the cavities are coupled. This happens as a result of reaching the strong coupling regime where the optomechanical coupling strength becomes comparable with $\kappa_1$ as shown in Fig. \ref{fig:P_b}.\par
\begin{figure}
\subfloat[\label{fig:R_a}]{\includegraphics[width=0.4\textwidth]{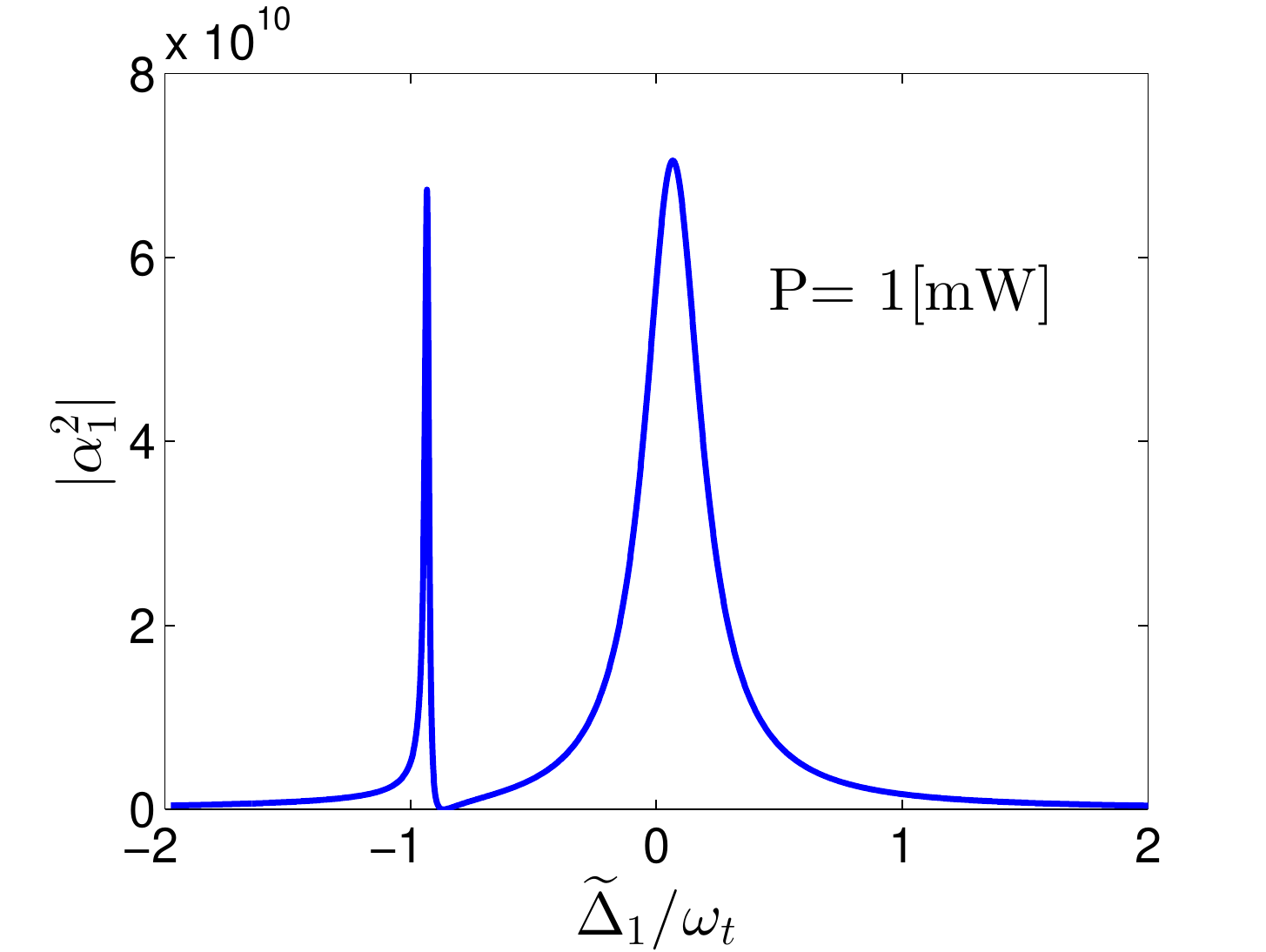}}
\subfloat[\label{fig:R_b}]{\includegraphics[width=0.4\textwidth]{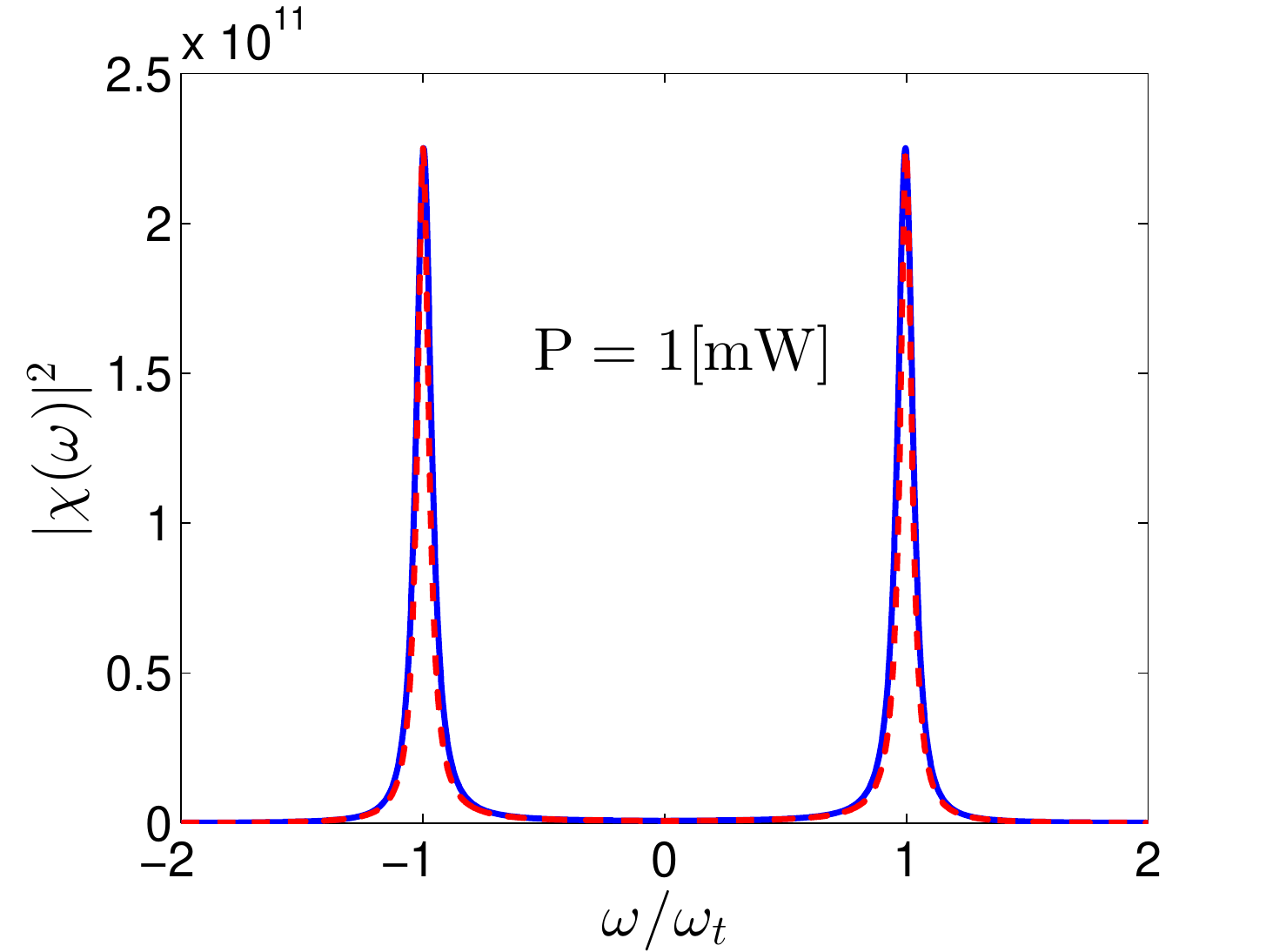}}\\
\subfloat[\label{fig:R_c}]{\includegraphics[width=0.4\textwidth]{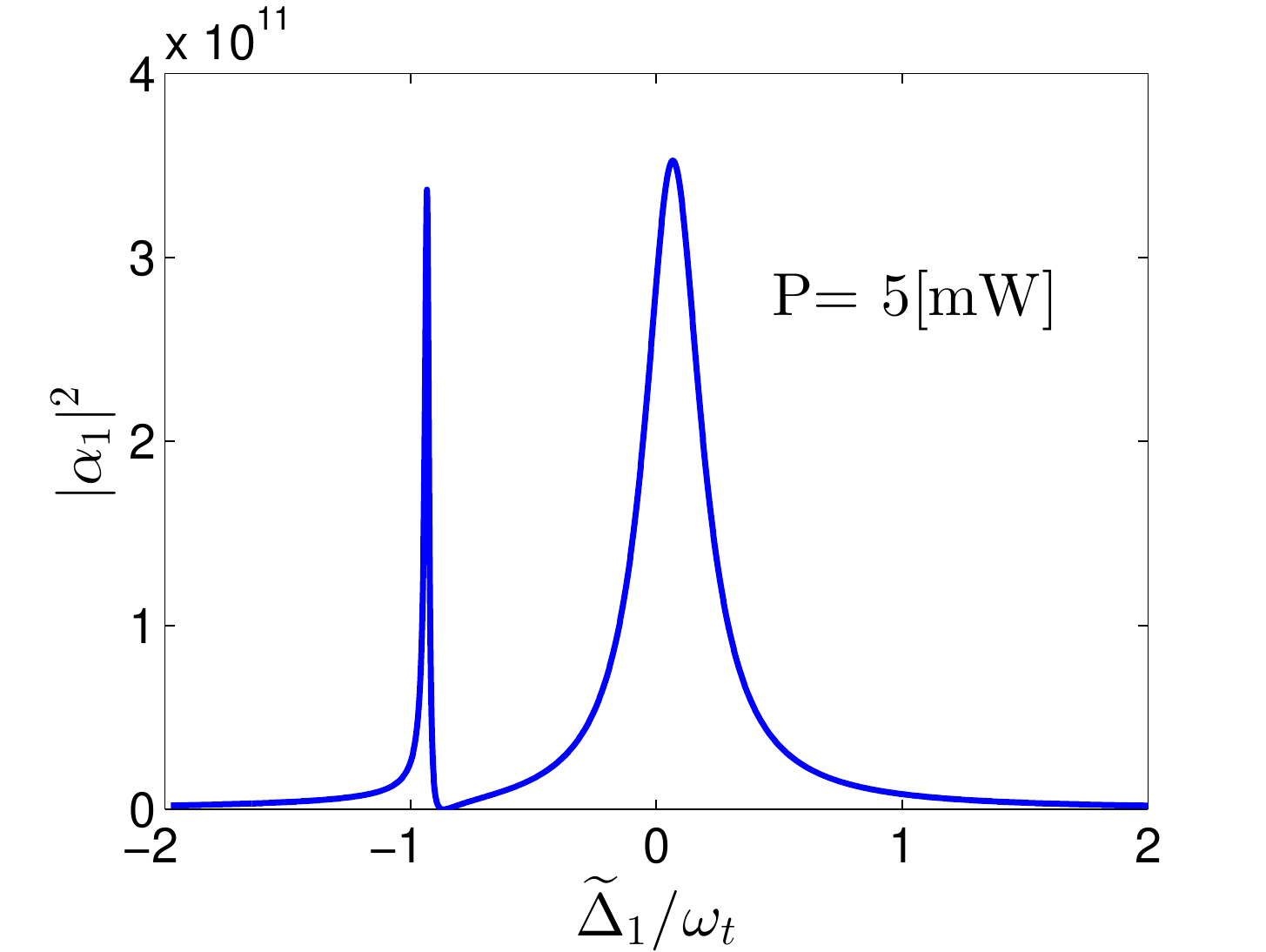}}
\subfloat[\label{fig:R_d}]{\includegraphics[width=0.4\textwidth]{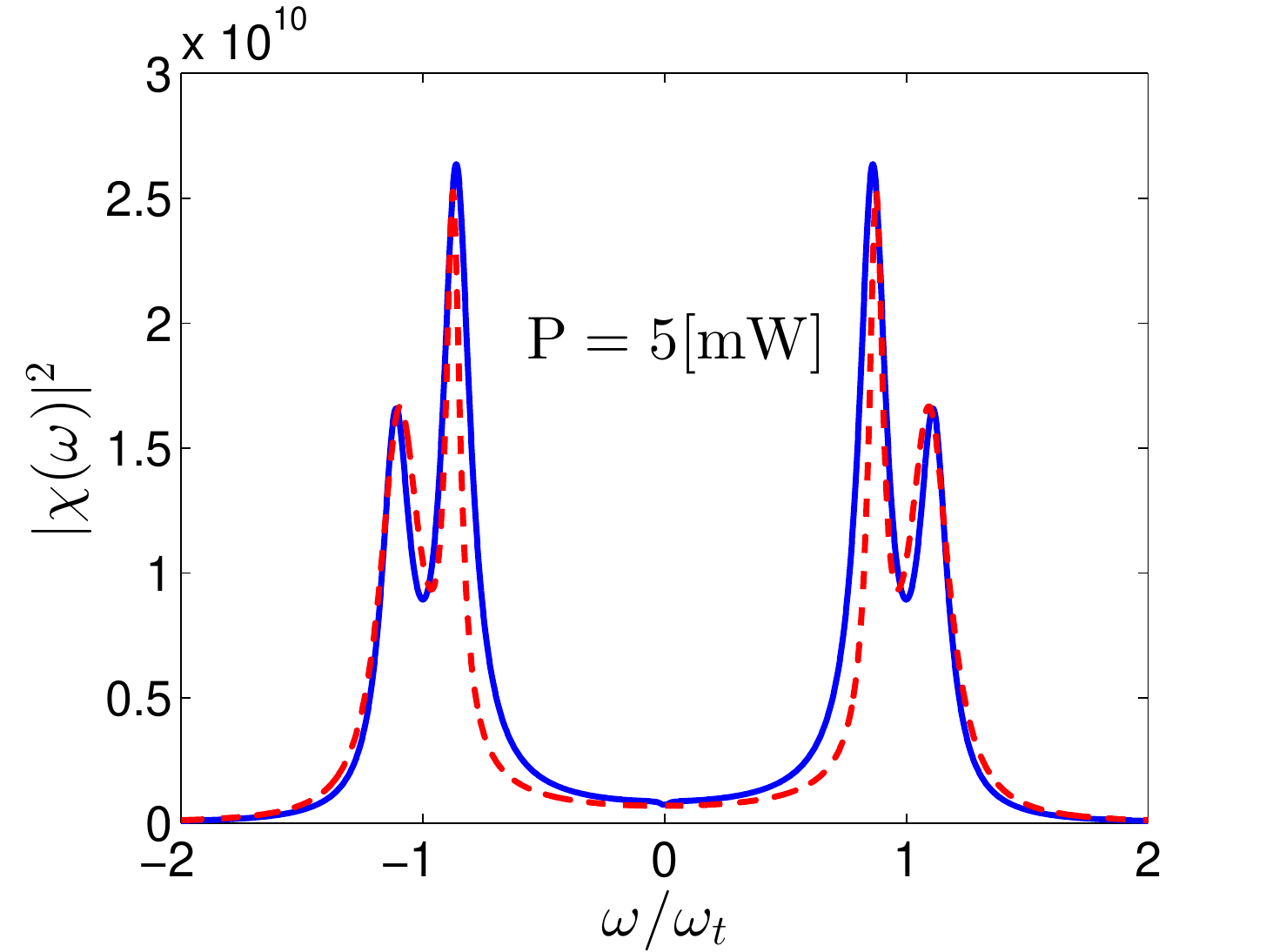}}
\caption{\label{fig:R}(a),(c) The number of photons inside the first cavity. (b),(d) The mechanical response of the system. The dashed lines show the approximated lineshapes correspond to Eq. (\ref{eq:SxxSL}) and Eq. (\ref{eq:SxxDL}).}
\end{figure}

Eventually, Fig. \ref{fig:R} shows the mechanical response of the system as well as the steady state number of photons inside the first cavity ($|\alpha_1|^2$) at two different laser powers. According to this figure, the mechanical response of the system has a Lorentzian lineshape at low laser powers, while it is made of two distinct resonances that look likes a double Lorentzian lineshape at higher ones. Furthermore, the steady state number of photons inside the first cavity shows the typical Fano resonance behavior at each laser powers. We have used this resonance to enhance the optomechanical coupling strength and consequently the cooling efficiency.
\section{\label{sec:C}Conclusion}
In this work, we have shown that the efficiency of optical cooling can be enhanced in a system of coupled cavities in the resolved sideband regime thanks to the presence of an extra Fano lineshape in the optical response of the system. The presence of the second cavity enables us to design the system in such a fashion that the steady state number of photons and consequently the optomechanical coupling parameter are maximized at the optimum detuning for cooling.\par
We have also obtained two closed solutions for the cooling rates by approximating the mechanical response of the system with either a single or double Lorentzian lineshapes. According to the single Lorentzian approximation, the optimum cooling is obtained when $\omega_m=\sqrt{d^2+4\mu^2}$ and $\widetilde{\Delta}_1=-\frac{\omega_m+d}{2}$. The results obtained from the double Lorentzian approximation always comply with the exact solutions even in the strong coupling regime. However, the single Lorentzian approximation fails to provide the exact values for the cooling rates in the strong coupling regime. Nevertheless, it has the benefit of predicting the optimum values of $\widetilde{\Delta}_1$ properly even in the strong coupling regime.
\bibliography{refrences}

\end{document}